\documentclass[preprint, showpacs,amsmath,amssymb,aps,prb]{revtex4}
\usepackage{graphicx}
\usepackage{color}
\usepackage[colorlinks,citecolor=blue,urlcolor=blue,bookmarks=false,hypertexnames=true]{hyperref}

\begin{document}


\title{Quantum point contact: a case of spin-resolved electron transport}

\author{Artur\ Useinov$^{1,2}$\footnote{artu@nctu.edu.tw, corresponding author*}, Hsiu-Hau\ Lin$^{3}$, Niazbeck Useinov$^2$ and Lenar\ Tagirov$^{2,4}$\footnote{ltagirov@mail.ru}}
\affiliation{$^1$International College of Semiconductor Technology, National Chiao Tung University, Hsinchu 300, Taiwan}
\affiliation{$^2$Institute of Physics, Kazan Federal University, Kazan, 420008, Russia}
\affiliation{$^3$Department of Physics, National Tsing Hua University, Hsinchu, 30013, Taiwan}
\affiliation{$^4$Kazan E. K. Zavoisky Physical -Technical Institute, Kazan, 420029, Russia}

\date{\today}

\begin{abstract}
The work presents the extended theoretical model of the electrical conductance in non-magnetic and magnetic nano-size point contacts. The developed approach describes diffusive, quasi-ballistic, ballistic and quantum regimes of the spin-resolved conductance. It is based on the electron transport through metallic junction within approach of the circular constriction. The model provides unified description of the contact resistance from Maxwell diffusive through the ballistic to purely quantum transport regimes without any residual terms depending from the scale. An application are given for experimental treatment of the golden nanocontacts as a non-magnetic case and Permalloy nanowire with/without domain wall as example for the magnetic system. The model of quantum point contact assumes that contact area can be replaced by the quantum object ({\it{i.e.}} magnetic tunnel junction, narrow domain wall, etc.), where potential energy profile determine its electrical properties.
\\

\end{abstract}

\maketitle

\section{\label{sec:level1}Introduction}

Quantitative theory of conductance $\sigma$ of the systems in confined geometry, such as point-like contacts, has great implications for determining the contact size \cite{Marrows,Naidyuk,Garcia1,Agrt,Tim4}. A simplest, but relevant in most cases solvable model for a point contact is a circular constriction of the radius $a$ which connects two large electron reservoirs. It is convenient to quantify the conducting properties of the nanocontact {\it{via}} the dimension ratio of the geometrical size $a$ to the electron mean free path $l$. The $a/l$ or  Knudsen number $K=l/a$ values come as an output of fitting a theory to experimental data on resistance of point nanocontacts \cite{Wex6}. Once the mean free path $l$ is known from resistivity measurements of the material, the effective diameter $d=2a$ of the contact can be estimated from the fitted $K$.

The model diameter $d$ can be identified as the size of the contact if information about the contact shape is unavailable. Two limiting regimes of conductance through nanocontacts are commonly discussed: the Maxwell, or diffusive conductance $\sigma _{M}$, when the contact size much larger than $l$ ($K \ll 1.0$) \cite{Max1,Hol3,Doudin}
\begin{equation}
\label{e1}
 \sigma _{M}  = 2a/\rho _{\rm{V}},
\end{equation}
where $\rho _{\rm{V}}$ is bulk resistivity, which can be expressed in terms of bulk conductivity $\sigma _{\rm{V}}$ of the isotropic metal as follows:
\begin{equation}
\label{e2}
\rho _{\rm{V}}^{ - 1}  = \sigma _{\rm{V}}  = \frac{{e^2 n\,l}}{{ \hbar k_F }}\, = \frac{{e^2 p_F^2 l}}{{3\pi ^2 \hbar ^3 }},
\end{equation}
where $e$, $ k_F = p_F/ \hbar$ and $n = k_F^3 /3\pi ^2 $ - electron charge, Fermi wave-number and free electron density in metals, respectively. Within the model, the bulk mean free path $l = \hbar k_F \tau /m_e$ depends on impurities, defects, electron-electron and electron-phonon scattering {\it{via}} averaged time between collisions, $\tau$ ($m_e$ is electron mass).

The second regime refers to the ballistic conductance through the contact area when no collisions occur during transmission of an electron through the constriction \cite{Sha7}, $K \gg 1.0$. If no collisions crossing the contact area, then no information about $l$ is present in the Sharvin conductance of the nanocontact:
 \begin{equation}
\label{e3}
\sigma _S  = \frac{{e^2 a^2 k_F^2 }}{{4\pi \hbar}} = G_0 N.
\end{equation}

The factor  $G_0 = 2e^2 /h $ is conductance quantum \cite{Doudin}, $N \simeq \left( {k_F a/2} \right)^2 $ is the number of opened conductance channels accommodating the nanoconstriction\cite{Lan8}.

Furthermore, it is relatively easy to obtain the expression which shows connection between $\sigma_S$, $\sigma _{\rm{V}} $ and $\sigma _M $: $\sigma _S  = \frac{{3\pi a}}{{4K}}\sigma _{\rm{V}}$, $\sigma _{\rm{V}}  = \frac{{4K}}{{3\pi a}}\sigma _S$ and $\sigma _M = \frac{{8\,K}}{{3\pi }}\sigma _S $. It is noticed that Sharvin {\it{et al.}} \cite{Sha7} estimated asymptotic behavior of resistivity as $R_S  \simeq p_F /e^2 \left( {2a} \right)^2 n$. An expression of Sharvin conductance in the form of $\sigma _S  = 3\pi /\left( {16R_S } \right)$ with accuracy up to $3\pi /16$ factor is used in literature \cite{Tim4, Ert9, Bra10, Gat11}. In general case $n$ is a complicated function of $k_F$. Hence, if the system is not limited by the model of the free electrons, the $n$ might be corrected based on the properties of the specified material and its Fermi surface. As a result, $n$ and $k_F$ can be determined in the range of {\it{ab-initio}} calculations.

Moreover, there is another form of Sharvin conductance which is often used \cite{Tim4, Ert9, Bra10, Gat11}:

\begin{equation}
\label{e4}
\sigma _{_S }^*  = \frac{{3\pi a^2 }}{{4\rho _{\rm{V}} l}}
\end{equation}

It is obtained by multiplying the numerator and denominator of (\ref{e3}) by $l$ and applying the dimension of $\rho _{\rm{V}}$, (\ref{e2}). Expression (\ref{e4}) has an advantage that in the case of constriction (or a contact of identical metals), it can be applied for estimation of the constriction radius $a$. Indeed according to (\ref{e2}), the product $\rho _{\rm{V}} l ={\rm{constant}}$ is independent of the mean free path $l$. Provided that $\rho _{\rm{V}} l$ product is known ($\rho _{\rm{V}}$ can be obtained from resistivity measurements, and $l$ can be extracted {\it{via}} size effects in thin films, or combining the resistivity with specific heat measurements and utilizing the Pippard relations \cite{Pippard1,Pippard2}), it seems, that (\ref{e4}) represents a useful tool to estimate the nanocontact size.

The problem can be considered from the opposite point of view: once the contact size is known in some way, the resistance measurements give a tool to estimate $\rho _{\rm{V}} l$ - product, i.e. establish the contact material parameter from the single kind of measurements. Indeed, it has been done in [\cite{Ert9}], where it was pointed out that next step to extract the mean free path $l$ from that $\rho _{\rm{V}} l$ = constant, has yielding $l = 3.8$\,nm which is an order of magnitude below the bulk $l \simeq 37$\,nm for 99.99\% pure gold at room temperature. Moreover, it was noticed that the range of applicability of the ballistic Sharvin approach, (\ref{e3}) or (\ref{e4}), is restricted to smallest radius of the contacts close to 1\,nm, otherwise, the accordance of the theory and the experiment is poor. Thus, the both diffusive Maxwell and ballistic Sharvin limits of the nanocontact conductance cover extreme limits keeping unexplored a wide gap of most accessible and relevant sizes from 1\,nm to 100\,nm.

The analysis of the electron transport through a circular constriction at arbitrary relationship between the orifice radius $a$ and the mean free path $l$ has been made by Wexler \cite{Wex6}. It was based on variation solution for the Green function (GF) of the Boltzmann kinetic equations. The solution obtained for the resistance was represented as follows \cite{Wex6}:

\begin{equation}
\label{e5}
R_{{\mathop{\rm W}} } =\frac{1}{{\sigma _W }}  = \frac{1}{{\sigma _M }}\gamma \left( K \right) + \frac{1}{{\sigma _S }},
\end{equation}
where $\sigma _W$ is determined as Wexler conductance, $\gamma (K)$ is a slowly varying function with the asymmetric values  $\gamma (K \to 0 ) = 1.0$, and
$\gamma (K \to \infty ) = 9\pi^2/128 = 0.694$. Expression (\ref{e5}) has a form of interpolation formula combining additively the diffusive Maxwell and ballistic Sharvin resistances with each of the counterparts vanishingly small when one of them approaches the related limit.

In 1999 Nikolic and Allen \cite{Nik20} improved accuracy of the Wexler solution for the orifice conduction, which is valid for only non-magnetic junctions. The stationary Boltzmann equation was solved taking into account Bloch-wave propagation and Fermi-Dirac statistics in presence of an electric field and Poisson equation for the electric potential. It is worth to note, that this solution is mentioned in literature as exact solution \cite{Tsym}. The result was compared with Wexler solution, where it is possible to re-define $\gamma (K)$ showing a corrected behavior. At the same time, Mikrajuddin {\it{et al.}} \cite{Mik21} proposed their approach to calculate orifice constriction resistance based on the solution of the electrostatic Laplace problem summing up the resistances of the infinitesimal shells between equipotential surfaces. The result is allowed to represent in the similar form as (\ref{e5}) with re-defined $\gamma (K)$. The comparison of the both Nikolic and Mikrajuddin solutions shows the valuable difference between them, that again induces the interest to the problem. To summarize, the calculation of the resistance in orifice constriction {\it{via}} the classical elecrodynamical approaches results in the sum of the diffusive and ballistic terms with problematic transition from one limit to another.

We propose an alternative approach which based on quasi-classical transport formalism \cite{Tag5}. The advantage of our solution is a simple integral expression which provides smooth functional transition between Sharvin and Maxwell limits without residual terms or counterparts. Moreover, the obtained result is derived as a limit case from general quantum model of the nanocontact constructed from different magnetic metals. As a result, the model fits experimental data for golden (symmetric, non-mangetic) contacts as well as simulates, for example, the resistance impact of the single domain wall in Permalloy (Py) nanowire.

\section{Model of the Magnetic Nanocontact}

In this section the magnetic contact was considered within quasi-classical approach which is based on GF formalism \cite{Tag5}. The application of this method is suitable for the structure dimensions larger than Fermi wave length of free electron, $\lambda _F  = 2\pi /k_F  \approx 0.5\;{\rm{nm}}$.

\begin{figure}
\includegraphics[width=0.85\textwidth]{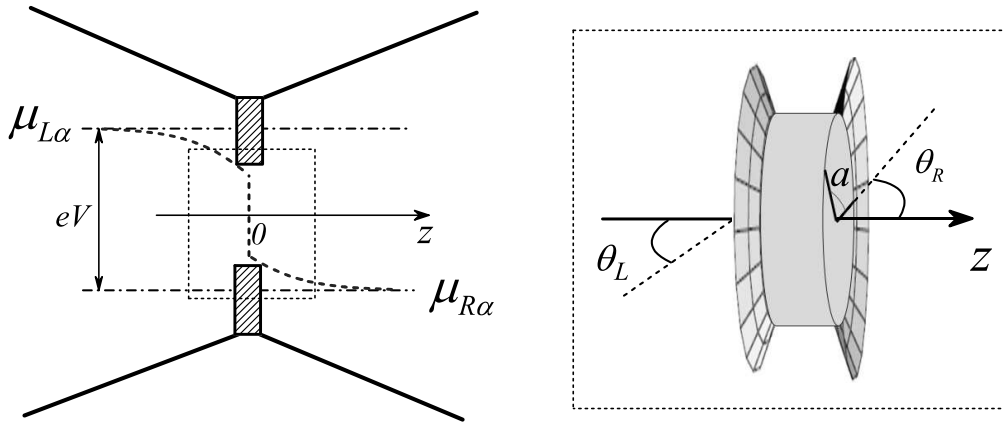}
\caption{The schematic view of the point contact with chemical potential drop. The selected rectangular area shows the contact area in non-conductive membrane. In general case, an electron with $p_{F,\alpha }^L$ and trajectory angle $\theta _{L,\alpha }$ transmits to the right-hand side of the contact, where different $p_{F,\alpha }^R$ and $\theta_{R,\alpha}$ are considered, respectively.}
\label{fig1}
\end{figure}

We consider the general case of ferromagnetic (FM) hetero-contacts, {\it{i.e.}}, the contact composed from different FM materials. Spin-dependent Fermi momenta $p_{F, \alpha}$ and electron mean free path $l_\alpha$ ($\alpha =  \uparrow , \downarrow $) in both contact sides are considered as arbitrary parameters. The point contact is modeled by a conductive spot (or a circular orifice) of the radius $a$ obtained in impenetrable membrane. This membrane divides the space into the left ($L$) and right ($R$) half-spaces, and each half-space is assigned to a single magnetic domain, Fig.\ref{fig1}.

The spherical coordinate system [${\bf{k}},\theta ,\phi $] is aligned with the cylindrical one [${\bf{r}} ,\phi ,z$], where the $z$ axis is perpendicular to the plane of the membrane. The voltage $V$, which induces the electrical current $I^z  = I_ \downarrow ^z  + I_ \uparrow ^z$, is applied far away from the contact area. It was demonstrated that the current density $j^z_\alpha$ with spin index $\alpha$ can be written as follows \cite{Tag5}:
\begin{equation}
\label{e6}
j_\alpha ^z (z \to 0,{\bf{k}},\varepsilon ) =  - \,\frac{{ep_{F,\alpha }^2 }}{{2\pi \,\hbar ^3 }}\left\langle {\cos \theta _\alpha  \,g_{a,\alpha } \left( {z,{\bf{k}},\varepsilon } \right)} \right\rangle,
\end{equation}
where $\varepsilon$, ${\bf{k}}$ and $\theta_{\alpha}$ - energy of electron, wave-number vector and the angle between $z$ axis and direction of the current flow, respectively. The angular brackets denote averaging over the solid angle, $\left\langle {(..)} \right\rangle  = \oint {(..)d\Omega \,} /2\pi$. The GF $g_a \left( {z,{\bf{k}},\varepsilon } \right)$ is Fourier transform of $ g_{a}(z,{\bf{r}},t)$ :
\begin{equation}
\label{e7}
g_{a(s)} \left( {z,{\bf{r}},t} \right) = \int {\frac{{d^2 k}}{{\left( {2\pi } \right)^2 }}\frac{{d\varepsilon }}{{2\pi }}g_{a(s)} \left( {z,{\bf{k}},\varepsilon } \right)} \,e^{ - i\left( {{\bf{k}}\,{\bf{r}} - \varepsilon \,t} \right)},
\end{equation}
where $g_{a(s),\;\alpha } \left( {z,{\bf{r}},t} \right) = {\textstyle{1 \over 2}}[g_\alpha  (n_z ,z,{\bf{r}}) \mp g_\alpha  ( - n_z ,z,{\bf{r}})]$ is quasi-classical GFs which is antisymmetric (symmetric) with respect to the projection $n_z  = p_{z\,,\alpha } /p_{F,\alpha } $ of the Fermi momentum on the $z$ axis, and ${\bf{r}}$ is radius-vector located in the contact plane.

The equation for the stationary charge current in a mixed representation {\it{via}} the current density (\ref{e6}), integrated over ${\bf{r}}$ in a cylindrical coordinate system at $z \to 0$, is as follows:
\begin{equation}
\label{e8}
I_\alpha ^z \left( {z \to 0} \right) = a\int_0^\infty  {dk\,J_1 \left( {ka} \right)j_\alpha ^z \left( {z,k} \right)},
\end{equation}
where $J_1 \left( x \right)$ is the Bessel function, appeared as a result of integration over the contact plane. The GFs $g_a \left( {z,{\bf{r}},t \to 0} \right)$ are solution of the kinetic equations along the quasi-classical trajectories, which were derived by Zaitsev \cite{Zai13} and adapted by Tagirov {\it{et al.}} \cite{Tag5,TagGar} in the form similar to (\ref{e9}). As a result, GFs can be applied to the description of the electron transport in classical and quantum nanocontacts:
\begin{equation}
\label{e9}
\begin{array}{l}
 l_{z,\alpha } \frac{{\partial g_{a,\alpha}(z,{\bf{r}}) }}{{\partial z}} + {\bf{l}}_{|| ,\alpha } \frac{{\partial g_{s,\alpha}(z,{\bf{r}}) }}{{\partial {\bf{r}}}} + g_{s,\alpha } (z,{\bf{r}}) - \left\langle {g_{s,\alpha }(z,{\bf{r}}) } \right\rangle  = 0, \\
 l_{z\,,\alpha } \frac{{\partial g_{s,\alpha }(z,{\bf{r}}) }}{{\partial z}} + {\bf{l}}_{|| ,\alpha } \frac{{\partial g_{a,\alpha }(z,{\bf{r}}) }}{{\partial {\bf{r}}}} + g_{a,\alpha }(z,{\bf{r}})  = 0, \\
 \end{array}
\end{equation}
where ${\bf{l}}_{||,\alpha }  = \tau {\bf{v}}_{||,\alpha }^{\rm{F}}$ is the vector direction which coincides with the transverse component of the Fermi velocity ${\bf{v}}_{||,\alpha }^{\rm{F}}$, parallel to the plane of the contact. The magnitude of the vector can be calculated as $l_{||,\,\alpha} = \sqrt {l_\alpha ^2  - l_{z,\alpha }^2 }$, where $l_{z,\alpha }  = l_\alpha  \cos \left( {\theta _\alpha  } \right) = v_z \tau$ is projection on the $z$-axis. The brackets determine an averaging over the solid angle $\left\langle {..} \right\rangle _{\theta _c }  = \frac{1}{{2\pi }}\int_0^{\pi /2} {\int_0^{2\pi } {(..)\,\,\sin \left( {\theta _c } \right)d\theta _c d\varphi }}$, where $c = (L,R)$ is index of the contact side.

Equations (\ref{e9}) need to satisfy the quantum boundary conditions (BCs) which were derived by Zaitsev\cite{Zai13} for the superconductive and normal metal junctions. The BCs were adapted further to FM contacts in [\cite{Tag5, TagGar}]:
\begin{equation}
\label{e10}
g_{a,L\alpha } (0, {\bf{r}}) = g_{a,R \alpha } \left(0,{\bf{r}} \right) = \left\{ {\begin{array}{*{20}c}
   {g_{a,\alpha } \left( { z=0, {\bf{r}}} \right),} & {k_{|| ,\alpha }  \le \min [ {k_{F,\,\alpha }^L ,\,k_{F,\,\alpha }^R } } ]  \\
   {0,} & {k_{|| ,\alpha }  > \min [ {k_{F,\,\alpha }^L ,k_{F,\,\alpha }^R } ]}  \\
\end{array}} \right.
\end{equation}
\begin{equation}
\label{e11}
2R_\alpha  g_{a,\alpha } \left(0, {\bf{r}} \right) = D_\alpha  \left\{ {g_{s,L \alpha } \left(0, {\bf{r}} \right) - g_{s,R \alpha } \left(0, {\bf{r}} \right)} \right\},
\end{equation}
where $k_{|| ,\alpha }$ is the projection of Fermi vector on the contact plane, $\min \left[ {k_1 ,k_2 } \right]$ means minimal by value between $k_1$ and $k_2$; $D_\alpha$ and $R_\alpha   = \left( {1 - D_\alpha  } \right)$ are the transmission and reflection coefficients, respectively. The related $D_\alpha$ is determined in terms of quantum mechanics and responsible for effects of the quantum interference in the area of the contact.

Chemical potentials difference between the contact sides $\left( {{\rm{\mu }}_{R \alpha}  - {\rm{\mu }}_{L\alpha} } \right)$ is proportional to the applied voltage potential $eV,$ as shown in Fig.\ref{fig1}. The solution for (\ref{e9}) can be simplified in the mixed coordinate and $k$-representation by performing the Fourier transform $g_{s(a),c} (z,{\bf{r}} ) = \left( {2\pi } \right)^{ - 2} \int {dk^2 } g_{s(a),c} (z,{\bf{k}})\;e^{ - i{\kern 1pt} {\bf{k}}{\bf{r}}}$:
\begin{equation}
\label{e12}
\begin{array}{l}
 \frac{{\partial ^2 g_s (z,{\bf{k}})}}{{\partial z^2 }} - \kappa ^2 g_s (z,{\bf{k}}) + \kappa l_z^{ - 1} \left\langle {g_s (z,{\bf{k}})} \right\rangle  = 0, \\
 \frac{{\partial  g_s (z,{\bf{k}})}}{{\partial z }} =  - \kappa g_a (z,{\bf{k}}), \\
 \end{array}
\end{equation}
where $\kappa  = \left[ {\left( {1 - i\,{\bf{kl}}} \right)} \right]l_z^{ - 1}$; the indexes $c$ and $\alpha$ are omitted. The homogeneous equation $\frac{{\partial ^2 g_s (z,{\bf{k}})}}{{\partial z^2 }} - \kappa ^2 g_s (z,{\bf{k}}) = 0$ has the following solution:
\begin{equation}
\label{e13}
g_s(z,{\bf{k}}) = C_1 e^{\kappa z}  + C_2 e^{ - \kappa z}  + C_0.
\end{equation}
The exact analytical form of the solution (\ref{e12}) was shown in [\cite{Tag5, Use14}] as follows:
\begin{equation}
\label{e14}
g_s (z,{\bf{k}}) = g_a (z,{\bf{k}})\,{\mathop{\rm sgn}}(z) + \frac{1}{{l_z }}\int\limits_{ - \infty }^{ + \infty }  \,e^{ - \kappa \left| {\xi  - z} \right|} \left\langle {g_s (\xi ,{\bf{k}})} \right\rangle d\xi.
\end{equation}
Equation (\ref{e14}) for the half-spaces can be represented in the forms:
\begin{equation}
\label{e15}
g_{sL} (z,{\bf{k}}) =  - g_{aL}(z,{\bf{k}})  + \frac{1}{{l_{zL} }}\int\limits_{ - \infty }^z  \,e^{ - \kappa _L (z - \xi )} \left\langle {g_{sL} (\xi ,{\bf{k}})} \right\rangle _{\theta _L } d\xi ,
\end{equation}
\begin{equation}
\label{e16}
g_{sR} (z,{\bf{k}}) = g_{aR}(z,{\bf{k}})  + \frac{1}{{l_{zR} }}\int\limits_z^\infty   \,e^{ - \kappa _R (\xi  - z)} \left\langle {g_{sR} (\xi ,{\bf{k}})} \right\rangle _{\theta _R } d\xi ,
\end{equation}
where $g_{aR} (z,{\bf{k}}) = g_a (z > 0,{\bf{k}})$ and $g_{aL} (z,{\bf{k}}) = g_a (z < 0,{\bf{k}})$ were used as related definitions.
One can consider the substitution $\eta  = \xi  - z$ and $z \to 0$ for (\ref{e16}):
\begin{equation}
\label{e17}
g_{sR} (0, {\bf{k}}) = g_{aR} (z,{\bf{k}})  + \frac{1}{{l_{zR} }}\int\limits_0^\infty   \,e^{ - \kappa _R \eta } \left\langle {g_{sR} (\eta, {\bf{k}})} \right\rangle _{\theta _R } d\eta.
\end{equation}
An approximate view of the integral distribution $\left\langle {g_{sR} (\eta,{\bf{k}})} \right\rangle _{\theta _R }$ can be considered in the form as (\ref{e18}) below, which is crucial in relation to the previous approaches, considered in Ref.[\cite{Tag5, Use14,NUse2015}] (see also Appendix\,A for details):
\begin{equation}
\label{e18}
\left\langle {g_{sR} (\eta, {\bf{k}})} \right\rangle _{\theta _R }  \approx \tilde c_R \,\eta \kappa _R \left\langle {g_{sR} (0, {\bf{k}})} \right\rangle _{\theta _R }.
\end{equation}

The numerical factor $\tilde c_R$ serves later as the calibration in the convergence of the final solution to the ballistic Sharvin and diffusive Maxwell limits without residual terms. Presented approximation (\ref{e18}) generates more accurate final result in comparison to the previous theoretical developments \cite{Tag5, Use14,NUse2015}. It is important to note that $\eta$ and $\kappa _R $ appear here symmetrically as a factors of the linearly behaved symmetric GF at $z=0$.

The resulted expression (\ref{e18}) was substituted into (\ref{e17}), the term $\left\langle {g_{sR} (0,{\bf{k}})} \right\rangle _{\theta _R } $ can be moved outside of the integral sign, since $\left\langle {g_{sR} (0,{\bf{k}})} \right\rangle _{\theta _R }  \approx {\rm{constant}}:$
\begin{equation}
\label{e19}
g_{sR} (0,{\bf{k}}) = g_{aR}(0,{\bf{k}})  + \tilde c_R l_{zR}^{ - 1} \left\langle {g_{sR} (0, {\bf{k}})} \right\rangle _{\theta _R } \int\limits_0^\infty   \,\eta \kappa _R e^{ - \kappa _R \eta } d\eta .
\end{equation}
A similar manipulation with (\ref{e15}) resulted in another equation for the left side of the contact:
\begin{equation}
\label{e20}
g_{sL} (0,{\bf{k}}) =  - g_{aL}(0,{\bf{k}})  + \tilde c_L l_{zL}^{ - 1} \left\langle {g_{sL} (0, {\bf{k}})} \right\rangle _{\theta _L } \int\limits_0^\infty   \,\eta \kappa _L e^{ - \kappa _L \eta } d\eta .
\end{equation}
Green function $g_s (z, {\bf{k}})$ is defined with accuracy up to the constant $C_0$, (\ref{e13}). This constant is equal to the equilibrium GF, $C_{0}  = g_s^{eq} (\varepsilon ) = {\rm{tanh}}({\textstyle{\varepsilon  \over {2k_{\rm{B}} T}}})$, where $\varepsilon$ is electron energy, and thus $g_s (z,{\bf{k}},\varepsilon )$ has to be redefined:
\begin{equation}
\label{e21}
g_s(z,{\textbf{k}},\varepsilon ) \equiv  f_s (z,{\textbf{k}}) + \,g_s^{eq} (\varepsilon )\Gamma(k),
\end{equation}
where $\Gamma(k)  = \int\limits_0^a dr \int\limits_0^{2\pi } r e^{(i \cdot {\bf{k}}{\bf{r}})} \,d\phi  = \frac{{2\pi a}}{k}\,J_1 (ka)$ is obtained due to integration of the $C_0$. As a result, (\ref{e15}) and (\ref{e16}) can be re-written after the substitutions $\eta  = \xi  - z$ at $z \to 0$:
\begin{equation}
\label{e22}
f_{s,c} (z,{\textbf{k}})  = {\rm{tanh}}\left( {\frac{\varepsilon }{{2k_{\rm{B}} T}}} \right)\Gamma (k)  \pm g_{a,c} (z,{\textbf{k}},\varepsilon)  + \mathop {\left\langle {f_{s,c} (z,{\textbf{k}}) } \right\rangle }\nolimits_{\theta _c } \tilde c_c \int\limits_0^\infty   l_{z,c}^{ - 1} \,\kappa _c \eta e^{ - \kappa _c \eta } \,d\eta,
\end{equation}
 where lower sign ($-$) have to be used for the left junction side, $c=L$. The approximation (\ref{e18}) has to be also redefined in the form $\left\langle {f_{s,c} (\eta ,{\bf{k}})} \right\rangle _{\theta _c }  \approx \tilde c_c \,\eta \kappa _c \left\langle {f_{s,c} (0,{\bf{k}})} \right\rangle _{\theta _c }$, where $\kappa_c = 1/\left[ {l_c x_c} \right]$. The system of the equations which defines $\left\langle {f_{s,c} } \right\rangle _{\theta _c ,\phi }$ is:

\begin{equation}
\label{e23}
\mathop {\left\langle {f_{sL}} \right\rangle }\nolimits_{\theta _L ,\phi }  =  - \mathop {\left\langle {g_{aL} } \right\rangle }\nolimits_{\theta _L ,\phi }  + \left\langle {f_{sL} } \right\rangle _{\theta _L ,\phi } \tilde c_L \int\limits_0^\infty   \mathop {\left\langle {{{\eta \kappa _L e^{ - \kappa _L \eta } } \mathord{\left/
 {\vphantom {{\eta \kappa _L e^{ - \kappa _L \eta } } {l_{zL} }}} \right.
 \kern-\nulldelimiterspace} {l_{zL} }}} \right\rangle }\nolimits_{\theta _L ,\phi } \,d\eta,
 \end{equation}

\begin{equation}
\label{e24}
\mathop {\left\langle {f_{sR}} \right\rangle }\nolimits_{\theta _R ,\phi }  = \mathop {\left\langle {g_{aR} } \right\rangle }\nolimits_{\theta _R ,\phi }  + \left\langle {f_{sR} } \right\rangle _{\theta _R ,\phi } \tilde c_R \int\limits_0^\infty   \mathop {\left\langle {{{\eta \kappa _R e^{ - \kappa _R \eta } } \mathord{\left/
 {\vphantom {{\eta \kappa _R e^{ - \kappa _R \eta } } {l_{zR} }}} \right.
 \kern-\nulldelimiterspace} {l_{zR} }}} \right\rangle }\nolimits_{\theta _R ,\phi } \,d\eta,
 \end{equation}
where $f_{sR} = f_{s,R} (z > 0,{\bf{k}})$, $f_{sL} = f_{s,L} (z < 0,{\bf{k}})$. The solution was found as follows:
 \begin{equation}
\label{e25}
\mathop {\left\langle {f_{s,c} (z,{\bf{k}})} \right\rangle }\nolimits_{\theta_{c} ,\phi }  = {\mathop{\rm sgn}\nolimits} (z)\mathop {\left\langle {g_{a,c}(z,{\bf{k}},\varepsilon) } \right\rangle }\nolimits_{\theta_{c},\phi } \left( {1 - \lambda_{c} } \right)^{-1},
\end{equation}
where $\lambda _{\,c}  = \tilde c_c \int\limits_0^\infty   \,\mathop {\left\langle {\left( {l_{z,c} } \right)^{ - 1} \eta \kappa _c \,e^{ - \kappa _c \eta } } \right\rangle }\nolimits_{\theta _c ,\phi } d\eta  = \tilde c_c \left( {1 + k^2 l_c^2 } \right)^{ - 1}$, (see Appendix B for details). It should be noted, that derived $\lambda_c$ as function over $k$ is crucially different solution in comparison to our previous works [\cite{Tag5, Use14,NUse2015}], where ${\lambda _c  = \left( {kl_c } \right)^{ - 1} \arctan \left( {kl_c } \right)}$. Finally, the equations (\ref{e22}) and (\ref{e25}) were substituted into (\ref{e11}) in order to obtain $g_a (z,{\bf{k}},\varepsilon)$:
\begin{equation}
\label{e26}
\begin{array}{l}
 2g_a \left( {z,{\bf{k}},\varepsilon } \right) =  - D \times \left[ {{\rm{tanh}}\left( {\frac{\varepsilon }{{2k_{\rm{B}} T}}} \right) - {\rm{tanh}}\left( {\frac{{\varepsilon  - eV}}{{2k_{\rm{B}} T}}} \right)} \right]\Gamma (k)  -  \\
  - \frac{{\mathop {\left\langle {g_{aL} } \right\rangle }\nolimits_{\theta _L } }}{{1 - \lambda _L }}\int\limits_0^\infty   D \times \left( {l_L x_L } \right)^{ - 2} \eta e^{ - \kappa _L \eta } d\eta  - \frac{{\mathop {\left\langle {g_{aR} } \right\rangle }\nolimits_{\theta _R } }}{{1 - \lambda _R }}\int\limits_0^\infty   D \times \left( {l_R x_R } \right)^{ - 2} \eta e^{ - \kappa _R \eta } d\eta,  \\
 \end{array}
\end{equation}
where $x_c  = \cos (\theta _c )$ and spin index $\alpha$ is omitted.

The averaging over the left and then over the right solid angles were applied for the both sides of the (\ref{e26}) to obtain the system with two equations. The unknown variables $\mathop {\left\langle {g_{aL} } \right\rangle }\nolimits_{\theta _L }$ and $\mathop {\left\langle {g_{aR} } \right\rangle }\nolimits_{\theta _R }$  where found considering BCs (\ref{e10}). Finally, the solution for $\mathop {\left\langle {g_{aL} } \right\rangle }\nolimits_{\theta _L }$ and $\mathop {\left\langle {g_{aR} } \right\rangle }\nolimits_{\theta _R }$ was return back into (\ref{e26}). The derived $g_a (z,{\bf{k}},\varepsilon)$  was substituted into equation for the current density (\ref{e6}). As a result, the final integration over $\varepsilon$ allows to find a spin-resolved current (\ref{e8}), where averaging over the solid angle of the left side was chosen for the convenience:
\begin{equation}
\label{e27}
I_\alpha ^{z}  = \frac{{e^2 p_{F\alpha }^2 a^2 V}}{{2\pi \hbar ^3 }}\int\limits_0^\infty   dk\frac{{J_1^2 (ka)}}{k}F_\alpha  (k),
\end{equation}
where $F_\alpha  (k) = \mathop {\left\langle {x_L D_\alpha  } \right\rangle }\nolimits_{\theta _L }  - \left( {G_1\mathop {\left\langle {x_L I_L } \right\rangle }\nolimits_{\theta _L }  + G_2 \mathop {\left\langle {x_L I_R } \right\rangle }\nolimits_{\theta _L } } \right)$, \\
$G_1 = \left\{ {\mathop {\left\langle {D_\alpha  } \right\rangle }\nolimits_{\theta _L } \left[ {2(1 - \lambda _R ) + \mathop {\lambda }\nolimits_2 } \right] - \mathop {\left\langle {D_\alpha  } \right\rangle }\nolimits_{\theta _R } \mathop {\lambda }\nolimits_4 } \right\}\Delta ^{ - 1}$, \\ $G_2 = \left\{ {\mathop {\left\langle {D_\alpha  } \right\rangle }\nolimits_{\theta _R } \left[ {2(1 - \lambda _L ) + \mathop {\lambda }\nolimits_1 } \right] - \mathop {\left\langle {D_\alpha  } \right\rangle }\nolimits_{\theta _L } \mathop {\lambda }\nolimits_3 } \right\}\Delta ^{ - 1}$, \\
$\Delta = 4\left( {1 - \lambda _L } \right)\left( {1 - \lambda _R } \right) + 2\left[ { \lambda _1 \left( {1 - \lambda _R } \right) +  \lambda _2 \left( {1 - \lambda _L } \right)} \right] -  \lambda _3  \lambda _4  +  \lambda _1  \lambda _2 $, where \\
  $\mathop {\lambda }\nolimits_1  = \int\limits_{\tilde x}^1 {dx_L } \frac{{D_\alpha  }}{{\left( {1 + \mathop {\left( {kl_L } \right)}\nolimits^2 \left( {1 - x_L^2 } \right)} \right)^{3/2} }}$, $\mathop {\lambda }\nolimits_2  = \int\limits_{\tilde x}^1 {dx_L } \frac{{\delta x_L D_\alpha  }}{{\sqrt {x_L^2  \pm x_{cr}^2 } \left( {1 + \mathop {\left( {kl_R \delta } \right)}\nolimits^2 \left( {1 - x_L^2 } \right)} \right)^{3/2} }}$, \\ $\mathop {\lambda }\nolimits_3  = \int\limits_{\tilde x}^1 {dx_L } \frac{{\delta \,x_L \,D_\alpha  }}{{\sqrt {x_L^2  \pm x_{cr}^2 } \left( {1 + \mathop {\left( {kl_L } \right)}\nolimits^2 \left( {1 - x_L^2 } \right)} \right)^{3/2} }}$, $\mathop {\lambda }\nolimits_4  = \int\limits_{\tilde x}^1 {dx_L } \frac{{D_\alpha  }}{{\left( {1 + \mathop {\left( {kl_R \delta } \right)}\nolimits^2 \left( {1 - x_L^2 } \right)} \right)^{3/2} }}$, \\
  $\mathop {\left\langle {x_L I_L } \right\rangle }\nolimits_{\theta _L }  = \int\limits_0^\infty   \,\mathop {\left\langle {x_L {\rm{ }}D_\alpha \frac{{\eta e^{ - \kappa _L \eta } }}{{l_{zL}^2 }}} \right\rangle }\nolimits_{\theta _L } d\eta  = \int\limits_{\tilde x}^1 {dx_L } \frac{{x_L D_\alpha  }}{{\left( {1 + \mathop {\left( {kl_L } \right)}\nolimits^2 \left( {1 - x_L^2 } \right)} \right)^{3/2} }}$, \\ ${\rm{ }}\mathop {\left\langle {x_L I_R } \right\rangle }\nolimits_{\theta _L }  = \int\limits_0^\infty   \mathop {\left\langle {x_L {\rm{ }}D_\alpha  \frac{{\eta e^{ - \kappa _R \eta } }}{{\left( {l_R x_R } \right)^2 }}} \right\rangle }\nolimits_{\theta _L } d\eta  = \int\limits_{\tilde x}^1 {dx_L } \frac{{x_L D_\alpha  }}{{\left( {1 + \mathop {\left( {kl_R \delta } \right)}\nolimits^2 \left( {1 - x_L^2 } \right)} \right)^{3/2} }}$. \\ \\
The boundary conditions (\ref{e10}) and (\ref{e11}) have to satisfy the conditions of the conservation, $k_{\parallel ,\alpha }  = k_{F,\alpha }^L \sin \left( {\theta _{L,\alpha } } \right) = k_{F,\alpha }^R \sin \left( {\theta _{R,\alpha } } \right)$, therefore, the lower integral limit $\tilde x$ is described as follows: $\tilde x = 0$ and $x_{cr}  = \sqrt {\left( {1 - \delta ^2 } \right)/\delta ^2 }$,  $\delta  = k_F^L /k_F^R$. The sign (+) should be taken in the root $\sqrt {x_L^2  \pm x_{cr}^2 }$, when electron moves from the state with  $k_F^L$ to $k_F^R$ and $k_F^L  < k_F^R$, otherwise, in the case $k_F^L  > k_F^R $, the sign is (-) as well as $\tilde x = x_{cr}  = \sqrt {1 - 1/\delta^{2} }$.

The solution (\ref{e27}) was verified for the non-magnetic limit case when $D_{ \uparrow , \downarrow }  = 1$, $l_{L,R}  = l$, $k_F^L  = k_F^R  = k_F $, $\lambda _{L(R)}  = \tilde c_{L(R)} \left( {1 + (kl)^2 } \right)^{ - 1}$.
Thus \\ $F_\alpha  (k) = \mathop {\left\langle {x_L } \right\rangle }\nolimits_{\theta _L }  - \left( {G_1 \mathop {\left\langle {x_L I_L } \right\rangle }\nolimits_{\theta _L }  + G_2 \mathop {\left\langle {x_L I_R } \right\rangle }\nolimits_{\theta _L } } \right) = \frac{1}{2} - \mathop {\left\langle {x_L I_L } \right\rangle }\nolimits_{\theta _L } \left( {G_1  + G_2 } \right)$, \\
where \\ $\mathop {\left\langle {x_L I_L } \right\rangle }\nolimits_{\theta _L }  = \mathop {\left\langle {x_L I_R } \right\rangle }\nolimits_{\theta _L }  = \int\limits_0^1 {dx\;} x\left[ {1 + \mathop {\left( {kl} \right)}\nolimits^2 \left( {1 - x^2 } \right)} \right]^{ - 3/2}  = \left[ {1 + k^2 l^2  + \sqrt {1 + k^2 l^2 } } \right]^{ - 1}$. \\
The complete solution for the conductivity is constructed as: \\
$\sigma  = \left( {I_ \uparrow ^z  + I_ \downarrow ^z } \right)/V = \frac{{e^2 p_{F \uparrow }^2 a^2 }}{{2\pi \hbar ^3 }}\int\limits_0^\infty   dk\frac{{J_1^2 (ka)}}{k}F_ \uparrow  (k) + \frac{{e^2 p_{F \downarrow }^2 a^2 }}{{2\pi \hbar ^3 }}\int\limits_0^\infty   dk\frac{{J_1^2 (ka)}}{k}F_ \downarrow  (k)$. \\
One can consider that $F_ \downarrow  (k) = F_ \uparrow  (k)$ for the non-magnetic contact, making the variable exchange $y = ka$ instead of $k$ and keeping in mind that $\int_0^\infty  {dy\;J_1^2 (y)/y}  = 1/2$, the final expression is combined as follows:\\
$\sigma  = 4\sigma _S \left( {\frac{1}{4} - \int\limits_0^\infty   \,\frac{{dy}}{y}\frac{{J_1^2 (y)}}{{1 + \mathop {\left( {y\,K} \right)}\nolimits^2  + \sqrt {1 + \left( {yK} \right)^2 } }}\left( {G_1  + G_2 } \right)} \right).$ \\
In order to connect this solution for exact Maxwell and Sharvin limit cases, it was found that necessary condition for $\tilde c_{L(R)}$ is satisfied to $\tilde c_{L(R)}  = 1.0$ resulting to $\left( {G_1  + G_2 } \right) = 1.0$ and providing final equation for the non-magnetic case. The derived analytical solution is used in the next section for experimental data fitting and comparison with another theoretical models.

One of the applications of the presented generalized approach of the quantum contact might be further development of the quantum integrated circuits \cite{Kasper} for the next generation of electronics below 16\, nm. Potentially, the model can deal with spin-resolved conducting properties of the nanoscale elements such as nano-sized contacts, large conductive molecules, tunnel junctions, quantum dots, quantum spin transistors as well as other devices with controlled interference of the electrons. As a first order of approximation, the contact area can be replaced by quantum object, its internal potential energy profile defines a key impact into electrical properties. For instance the spin-resolved quantum (tunneling) part $F_\alpha  (k) = \left\langle {x_L D_\alpha  } \right\rangle _{\theta _L }$ of the presented model was successfully applied for the simple MTJs \cite{Use15, Use16} as well as for MTJs with embedded nanoparticles \cite{Use17, UseS7} explaining voltage behavior of the tunnel magnetoresistance (TMR-$V$) and $R$-$V$ curves. It is possible to calculate tunneling current, {\it{e.g.}}, in FM/barrier/FM system for parallel (P) and antiparallel (AP) states, calculating ${\rm{TMR}} = \left( {I^{z({\rm{P)}}}  - I^{z({\rm{AP)}}} } \right)/I^{z({\rm{AP)}}}$, since the transmission $D$ fully determines the tunneling properties. The $k_{F,\alpha }^{L(R)}$ values can be considered as effective one for both FM/barrier interfaces \cite{UseS7}, $k_{F,\alpha }^{L(R)} \sim  3 - 12$ nm$^{-1}$. Their values might reflect the barrier properties to filter out an electron wave-functions: it is well known that MgO(100) barrier works as a filter for minority electron wave-functions with related symmetry in Co/MgO/Co and Fe/MgO/Fe junctions \cite{But18, Zha19}. For example, in case of FM/MgO it is assumed that $k_{F, \downarrow }^{L\left( R \right)}$ is smaller than for FM/Al$_2$O$_3$ interface, that increases the value of spin polarization $P$, where $P = \left( {k_{F, \uparrow }  - k_{F, \downarrow } } \right)/\left( {k_{F, \uparrow }  + k_{F, \downarrow } } \right)$.

\section{\label{sec:level2} Results and Discussions}

\subsection{Comparison with alternative theoretical models}

One of the goals for this work is to show also the classical conductance for the non-magnetic junction. The approach presented below includes exact Maxwell and Sharvin analytical limits in such terms that their transformations to each other are smoothly resolved. Moreover, non-magnetic limit of the extended quantum transport model was derived in the form which doesn't include $\gamma$ parameter:
\begin{equation}
\label{e28}
\sigma  = 4\,\sigma _S \left( {\frac{1}{4} - \int\limits_0^\infty   \,\frac{{dy}}{y}\frac{{J_1^2 (y)}}{{1 + \mathop {\left( {y\,K} \right)}\nolimits^2  + \sqrt {1 + \left( {y\,K} \right)^2 } }}} \right)
\end{equation}
At $K \to \infty $ ($a/l \to 0$), it is clearly seen that integral is equal to zero and, hence, the conductance transforms into the ballistic case, $\sigma  = \sigma _S $. At $K \to 0$ ($a/l \to \infty $) the integral has an exact asymptotic, $\mathop {\lim }\limits_{K \to 0} \,\,\int\limits_0^\infty   \,\frac{{dy}}{y}\frac{{J_1^2 (y)}}{{1 + \mathop {\left( {y\,K} \right)}\nolimits^2  + \sqrt {1 + \left( {y\,K} \right)^2 } }} = $ $\frac{1}{4} - \frac{2}{{3\pi \,}}K$, and then it gives accurate solution of the diffusive limit, $\sigma  = \left( {8/3\pi } \right)K \sigma _S = \sigma _{M}$. It was found that quasi-ballistic approach of our model is more accurate than commonly used Wexler solution. It is worth to note, the present approach is more accurate in terms of achieving an exact Maxwell limit and estimations of the spin-resolved diffusive conductance in comparison to our previous works\cite{Tag5,Use14,NUse2015}. Numerical solution (\ref{e28}) is applicable for $l$ and $k_F$ estimations in symmetric (homo) non-magnetic contacts, while common solution (\ref{e27}) is applied for magnetic and non-magnetic asymmetric (hetero) contacts as well.

One of the attempt to improve accuracy of the Wexler solution was suggested by Nikolic {\it{et\,al.}}\cite{Nik20}. The explicit form of the GF was found in range of semi-classical transport theory for the orifice geometry. However, possible effects of the quantum interference in the area of the contact, which might be similar to BCs (\ref{e10}) and (\ref{e11}), were neglected. It was found that the most complicated, but more accurate Nikolic solution ($\sigma _{W'}$) might be fitted by (\ref{e5}) with accuracy of $1.0\, \%$, where $\gamma _{{\rm{fit}}}  \approx (1 + 0.83K)/(1 + 1.33K)$, the details see in [\cite{Nik20}].

\begin{figure}
\includegraphics[width=0.99\textwidth,clip]{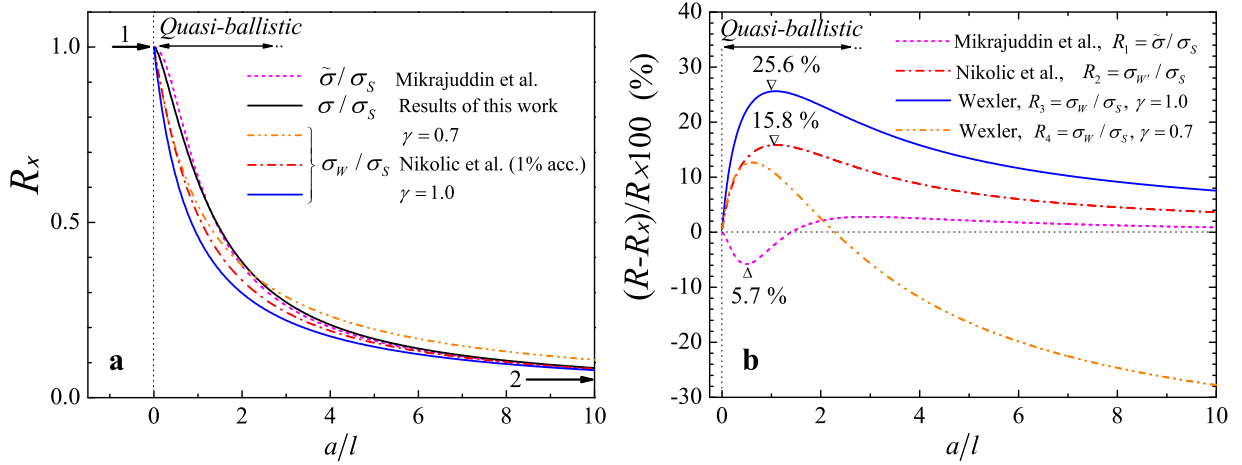}
\caption{ (a) The numerically estimated ratios between conductivity of the different models ($x = 1..4$) to the ballistic limit. Arrows 1 and 2 point to the ballistic and diffusive limits, respectively. (b) Relative difference of the related ratios to the solution of present work.}
\label{fig2}
\end{figure}

 The ratios of the conductance to Sharvin limit are collected in Fig.\ref{fig2}a for four model approaches, the result of the present work is shown as $\sigma /\sigma _S$. The ratios $\tilde \sigma /\sigma _S$, $\sigma _{W'} /\sigma _S$ and two cases of $\sigma_W /\sigma _S$ correspond to solution by Mikrajuddin {\it{et al.}}\cite{Mik21} with $\gamma  \approx {\textstyle{2 \over \pi }}\int_0^\infty  {e^{ - K \cdot x} {\rm{sinc}}(x)\;dx}$, solution by Nikolic with  $\gamma_{{\rm{fit}}}  \approx (a/l + 0.83)/(a/l + 1.33)$ and, finally, Wexler approaches with $\gamma  = 0.7$, $\gamma  = 1.0$, respectively. The relative comparison of the ratios, highlighted in Fig.\ref{fig2}b, shows that Mikrajuddin ratio is the closest one to our result, which is shown as zero level. Wexler-kind solutions show the difference of $11\, \%$ and $25.6\, \%$ for $\gamma=0.7 $ and $\gamma=1.0 $ at $a/l = 1.0$, respectively.

Many experimental works use some manipulations with $\gamma$ in (\ref{e5}) in order to achieve the desired fitting, since its value is commonly considered as a constant, $\gamma  \approx 0.7 - 0.75$, e.g. [\cite{Tim4, Ert9, Gat11}]. We assume that reduced value of $\gamma$ may lead to the compensation of an inaccuracy in Wexler model for the ballistic and quasi-ballistic regimes ($a \approx l$), e.g. when $a/l \to 0$ and $a/l \approx 2$, respectively, Fig.\ref{fig2}b. However, it might also lead to a large deviation of the estimated $l$ and $\sigma$ from real values using another scales. Hence, the value of $\gamma$ is considered to be a weak place of Wexler model.

\subsection{Conductance of the golden nanocontacts}

An experimental data following to Erts {\it{et al.}} \cite{Ert9} was considered for numerical comparison to our approach. The conductance for the point-like nanocontacts from gold (Au) was measured with different dimensions \cite{Ert9} and finally concluded that $l$ in nanocontact is supposed to decrease down to 3.8\,nm following to Wexler model even though $l$ in the bulk reaches 37\,nm. However, any explanation of this phenomenon was absent.

\begin{figure}
\includegraphics[width=1.1\textwidth]{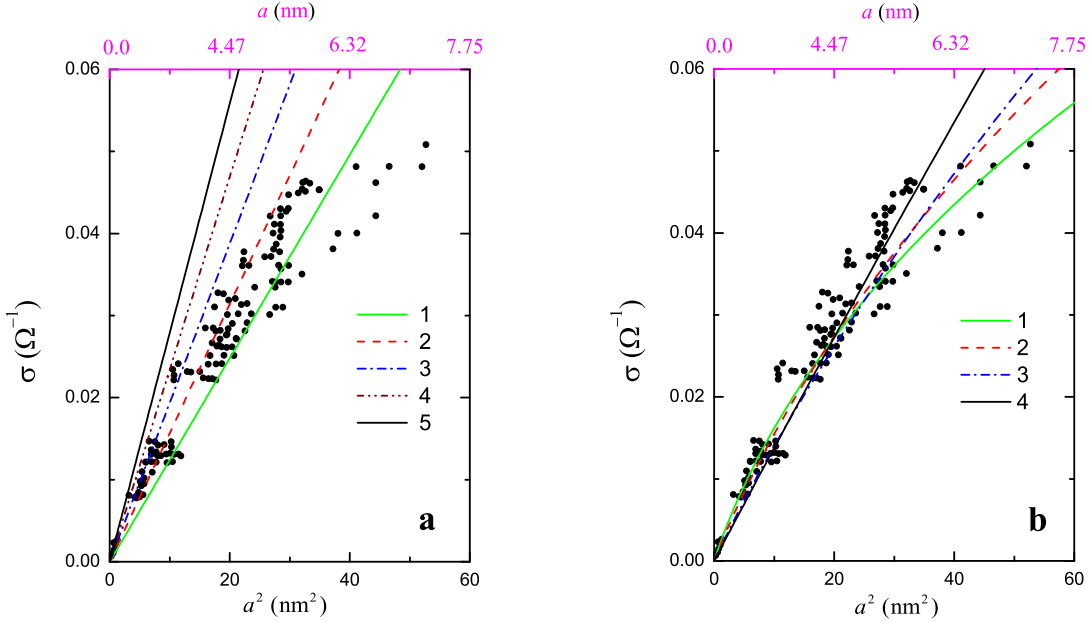}
\caption{Conductivity of the nanocontacts fabricated from gold with various contact dimensions. (a) Theoretical curves 1-5 are ascribed to the ballistic regime at $k_F^{{\rm{Au}}}  = 8.0\;{\rm{nm}}^{ - 1}$, $k_F^{{\rm{Au}}}  = 9.0\;{\rm{nm}}^{ - 1}$, $k_F^{{\rm{Au}}}  = 10.0\;{\rm{nm}}^{ - 1}$, $k_F^{{\rm{Au}}}  = 11.0\;{\rm{nm}}^{ - 1}$ and $k_F^{{\rm{Au}}}  = 12.0\;{\rm{nm}}^{ - 1}$, respectively. (b) The curves 1-4  are corresponded to $k_F^{{\rm{Au}}} = 11.0\;{\rm{nm}}^{ - 1}$, $k_F^{{\rm{Au}}}  = 10.0\;{\rm{nm}}^{ - 1}$, $k_F^{{\rm{Au}}}  = 9.0\;{\rm{nm}}^{ - 1}$, and $k_F^{{\rm{Au}}}  = 8.5\;{\rm{nm}}^{ - 1}$ with $ l = 4.0\;\rm{nm}$, $ l = 6.0\;\rm{nm}$, $ l = 10.0\;\rm{nm}$ and $ l = 38.0\;\rm{nm}$, respectively. Black dots are referred to the experimental data \cite{Ert9}.}
\label{fig3}
\end{figure}

Considering the ballistic conductivity of the golden contacts, we found that Sharvin limit at $k^{\rm{Au}}_F  = 8.0 - 12.0\,\,{\rm{nm}}^{ - 1} $ almost completely satisfies the experimental evidences presented in Ref.[\cite{Ert9}], Fig.\ref{fig3}a. It is easy to estimate the Fermi wavenumber in the bulk for Au applying $k^{\rm{Au}}_F=(3 \pi^2 n)^{1/3}$ with the following relations for $n$:
\begin{equation}
\label{e28dop}
n = v\rho N_A /A_r,
\end{equation}
where $v, \rho, N_A$ and $A_r$ - valency, density ($\rho^{\rm{Au}}=19.32$\,g/cm$^3$), Avogadro's number and relative atomic mass ($A^{\rm{Au}}_r= 196.967$). As a result, $k^{\rm{Au}}_F= 12.05$\,nm$^{-1}$ at $n \approx 5.9 \times 10^{22}$ and $k^{\rm{Au}}_F= 17.37$\,nm$^{-1}$ at $n \approx 1.77 \times 10^{23}$ are obtained in the cases of $v=1$ and $v=3$, respectively.

Nevertheless, one can partly analyze the possible cases when theoretical curves can show the best matching to the experimental data with the modified $k_F$ and $l$ due to some hidden reasons. The equation (\ref{e28}) was applied for the numerical estimation of the conductance in Au contact:
\begin{equation}
\label{e29}
\sigma ^{{\rm{Au}}}  = 4\sigma _S^{{\rm{Au}}} \left( {\frac{1}{4} - \int\limits_0^{Num}  \,\frac{{dy}}{y}\frac{{J_1^2 (y)}}{{1 + \mathop {\left( {y\,l/a} \right)}\nolimits^2  + \sqrt {1 + \left( {y\,l/a} \right)^2 } }}} \right)
\end{equation}
where $\sigma _S^{{\rm{Au}}}  = G_0 \left( {k_F^{{\rm{Au}}} a/2} \right)^2 $, $G_0  = 7.7481 \cdot \,10^{ - 5} \,\Omega ^{ - 1} $, upper limit of integral can be specified by $Num = 10^4$ for numerical estimation instead of $Num = \infty$. Figure \ref{fig3}b shows theoretical curves of the contact conductivity derived by (\ref{e29}), where $k_F^{{\rm{Au}}} $ and $l$ values were considered as independent parameters. Fitted curves 1 -- 3 and 4 are referred to $k_F^{{\rm{Au}}}  = 11.0\,{\rm{ nm}}^{ - 1} $, $k_F^{{\rm{Au}}}  = 10.0\,{\rm{nm}}^{ - 1} $, $k_F^{{\rm{Au}}}  = 9.0\,{\rm{nm}}^{ - 1}$ and $k_F^{{\rm{Au}}}  = 8.5\,{\rm{nm}}^{ - 1}$ with $l = 4.0\,{\rm{nm}}$, $l = 6.0\,{\rm{nm}}$, $l = 10.0\,{\rm{nm}}$ and $l = 38.0\,{\rm{nm}}$, respectively. Utilizing (\ref{e2}) in the form $n = k_F^{{\rm{Au}}} /\left( {\pi \,l\,G_0 \,\rho^{\rm{Au}}_{\rm{V}} } \right)$, where $\rho^{\rm{Au}}_{\rm{V}}=22.14\,\Omega  \cdot {\rm{nm}}$, the related parameters correspond to $n = 5.1 \cdot 10^{23} \,{\rm{cm}}^{ - 3} $, $n = 3.09 \cdot 10^{23} \,{\rm{cm}}^{ - 3}$, $n = 1.67 \cdot 10^{23} \,{\rm{cm}}^{ - 3}$ and $n = 4.15 \cdot 10^{22} \,{\rm{cm}}^{ - 3}$ for the curves 1 -- 4, respectively. Thus, the curves 3 and 4 have the closest values by $n$ with those which are estimated above for the $v=3$ and $v=1$, according (\ref{e28dop}). Finally, as a result, the most number of electrons, which passing through the contact area, probably keep the bulk mean free path. The conductance regime corresponds to the ballistic case, which is $l$-independent. More collected points are needed at $a > 6 \; {\rm{nm}}$ ($a^2  > 36\;{\rm{nm}}^{\rm{2}} $) in the range of quasi-ballistic and diffusive regimes to provide reasonable $l$-value estimations. Furthermore, the experimental data, showing the divergence to the curve 4 (Fig.\ref{fig3}b), e.g. the one around the curve 3, might correspond to Au (or Au-based defects) with $v=3$, otherwise anisotropic effects might also take place\cite{Doudin}.

\subsection{\label{sec:level2} Domain wall resistance in Py nanowire}

\begin{figure*}
\includegraphics[width=0.95\textwidth]{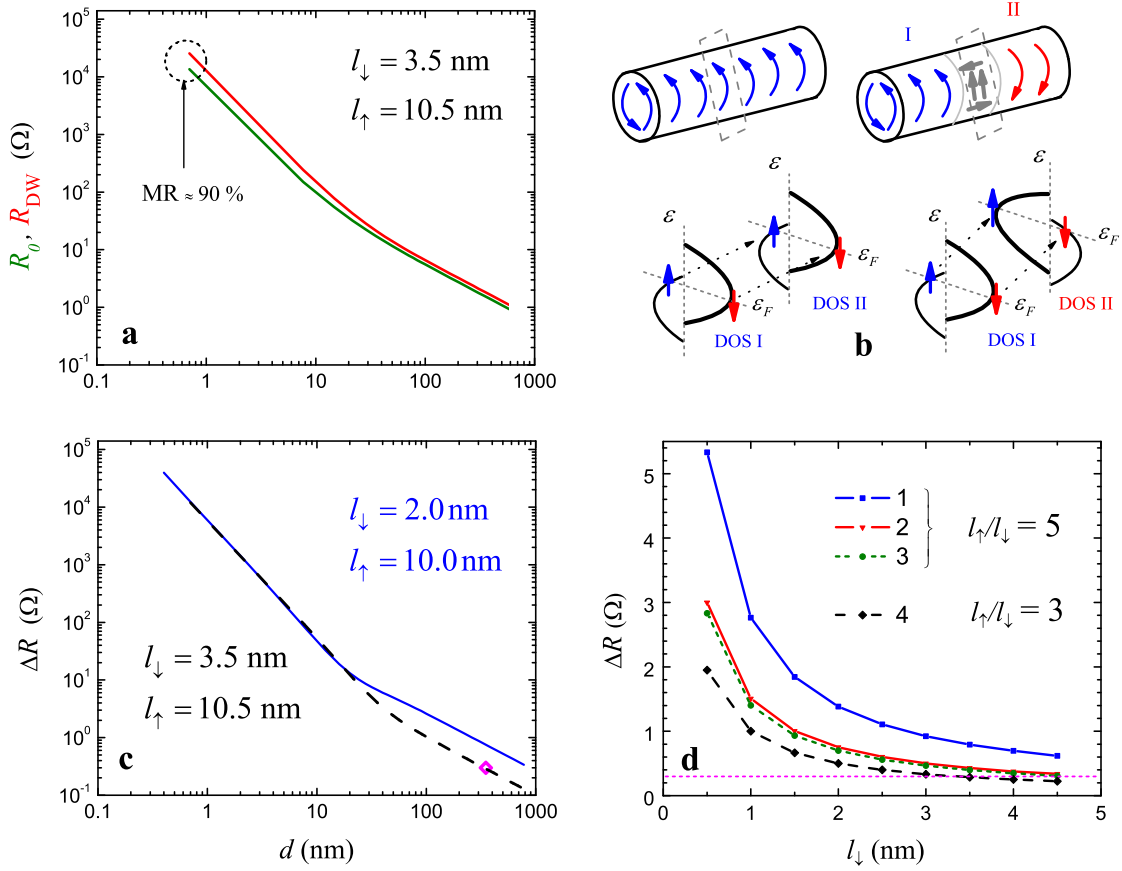}
\caption{(a) The red and green (bottom) solid curves show the resistance values of Py nanowire  $R_{{\rm{DW}}}$ ($R_{0}$) with and without DW, respectively. (b) The sketch of electron transitions and DOS differences at two vortex-like magnetic states: with DW (right) and without (left). (c) Clear impact of the DW resistance is shown, the dashed curve fits an experimental point (rhombic symbol) more precisely. (d) The results are demonstrated for DW resistance {\it vs.} $l_{\downarrow}$ derived for the different mean free path ratios and spin asymmetry of $k_{F \uparrow \downarrow}$-values.}
\label{fig4}
\end{figure*}

Finally, the resistance states were calculated in Py (Ni$_{80}$Fe$_{20}$) nanowire with and without single domain wall (DW) demonstrating the regime of the spin-splitted ballistic and diffusive electron transport (Fig.\ref{fig4}a). Utilizing the complete magnetic case of (\ref{e27}) and approach of the small voltage, when integral distribution is almost voltage independent, the resistance was estimated as $R=V/\left( {I_ \uparrow   + I_ \downarrow } \right)$. The area between vortex-like states was considered as one dimensional magnetic structure such as a narrow DW. Domain wall impact was integrated into the present model in similar way as in [\cite{Use14, Useinov}], where the transmission coefficient was considered as an exact analytical solution for the slope-like energy profile between shifted spin-resolved conduction band bottoms. The value of the spin-resolved density of states (DOS) at Fermi level is proportional to $k_{F, \alpha}$. The resistance becomes larger since the  electron scattering is enhanced in the DW transition area: spin diffusive length is assumed to be much larger than DW width (which taken as 3.0 nm in our case), and electron's spin is conserved over the transition while the definition of the spin subbands are opposite for the two magnetic parts of nanowire, Fig.\ref{fig4}b.

The theoretical estimations were compared with experimental measurement of the resistance difference, $\Delta R = R_{\rm{DW}}-R_{0}$ are taken from [\cite{Wong}], Fig.\ref{fig4}c and Fig.\ref{fig4}d. The difference arises between two opposite vortex-like magnetic states, Fig.\ref{fig4}b. Single experimental point $\Delta R \approx 0.3 \, \Omega$ is shown as a rhombic sign in Fig.\ref{fig4}c in Py nanowire with diameter $d=350$\,nm, and marked as horizontal dashed line in Fig.\ref{fig4}d.

 The presented $\Delta R(d)$ curves in Fig.\ref{fig4}c show the rapid drop-like behaviors which are changed in the region $d\approx 2\, l_ \uparrow$. The quasi-ballistic threshold is determined by the transition of the spin-up conductance from ballistic into diffusive regime. The complete diffusive regime is achieved for the both conductive spin channels when $d \gg l_{ \alpha }$. The black dashed curve is derived at $l_ \downarrow   = 3.5$\,nm and $l_ \uparrow   = 10.5$\,nm ($l_ \uparrow  /l_ \downarrow   = 3$) that allows to fit the experimental point $\Delta R = 0.3 \, \Omega$. The blue solid line, resulting in $\Delta R = 0.73\, \Omega$ at $d = 350$\,nm, is obtained for $l_ \downarrow   = 2.0$\,nm and $l_ \uparrow   = 10.0$\,nm; the considered $k_{F}^{\rm{Py}}$, $l_ \uparrow  /l_ \downarrow   = 5$ as well as $l_{\alpha}$ values itself, at this case, correspond to the realistic Py parameters. It is worth to notice, $\Delta R$ can be decreased in presence of the additional spin-flip leakages, which are not considered within our approach.

Experimental estimations of the spin-resolved $l$ are accessed in [\cite{Gurney, Nalwa}]. The considered $k$-values, which are attributed to Py nanowire, are also consistent with literature: $k_{F \uparrow \downarrow}^{\rm{Py}}$ are similar to Mu-metal (Py-type) compound \cite{Use14, Useinov}. Furthermore, ballistic approach of our model generates MR $\approx$\,90\,\% and 100\,\% for 3\,nm and 1\,nm of the DW width, respectively. These values are consistent with ballistic MR $\approx$\,100\% in Ni-Mu point-like junction measured experimentally \cite{Zhao} and estimated numerically \cite{Use14} at $d \approx 1$\,nm.

Figure\,4d shows additional $\Delta R$ simulations over $l_{\downarrow}$ taking into account the asymmetry difference between $k_{F \downarrow}^{L\left( R \right)}$ and $k_{F \uparrow}^{L\left( R \right)}$ for the fixed $k_{F \downarrow }^{L\left( R \right)}  = 10.8$ \ nm$^{-1}$ values. The curves 1 to 4 correspond to $k_{F \uparrow }^{L\left( R \right)}  = 4.1$\ nm$^{-1}$, $k_{F \uparrow }^{L\left( R \right)}  = 6.1$\ nm$^{-1}$, $k_{F \uparrow }^{L\left( R \right)}  = 8.1$\ nm$^{-1}$ and $k_{F \uparrow }^{L\left( R \right)}  = 6.1$\,nm$^{-1}$, respectively. Among the results the most rapid reduction of $\Delta R$ with $d$ corresponds to the smallest $l_ \uparrow  /l_ \downarrow$ and $\left|P\right|$ values.

\section{Conclusion}
In present work a transport model for magnetic and non-magnetic point-like contacts was reconsidered, making the approach more consistent. The final common solution allows to provide the smooth functional transition between Sharvin and Maxwell classical limits without rudiment terms. The possible replacement of the contact area by quantum object is one of the model's benefit. As a result, the spin-slitted potential energy profile of the quantum system defines a final behavior of the electrical properties. The conductance or resistance simulations in the nano and micro sized structures were shown utilizing the approaches of spin-resolved quantum, ballistic, quasi-ballistic and diffusive regimes. The developed theory has a wide spectra of applications from simple non-magnetic contacts to spin-resolved quantum limit such as the tunneling regime or simulation of spin-dependent quasi-ballistic conductance of the different nano-structures. For example, the conductance in Au nanocontacts has been calculated more accurately resulting in conclusion about ballistic behavior, in addition, the permalloy nanowire with and without DW was simulated in the large range of the dimensions showing the reasonable consistency of the DW resistance with experimental data. Non-magnetic limit of the developed approach for the metallic junctions demonstrates a valuable corrections with respect to another theoretical approaches presented in literature.

\section*{Acknowledgments}

We are thankful to Dr. Vitaly Gurylev (National Tsing Hua University) for a discussions and technical support. A. Useinov acknowledge partial support by the Program of Competitive Growth of Kazan Federal University. The reported study was also partially funded by RFBR according to the research project N$^{\b{o}}$ 18-02-00972.

\section*{Appendix A}
One can consider: $\left\langle {g_{sR} \left( {\eta ,{\bf{k}}} \right)} \right\rangle _R  = \tilde c_R \kappa _R \eta \left\langle {g_{sR} \left( {0,{\bf{k}}} \right)} \right\rangle _R$ for $z > 0$, $ z \to 0$ at $\eta  = \xi  - z$ and assuming $\kappa _R  \approx 1/l_{zR}  = \left( {lx} \right)^{ - 1}$, equation (\ref{e18}) can be obtained in the following form:
\begin{equation}
\label{A1}
\left\langle {g_{sR} \left( {\xi ,{\bf{k}}} \right)} \right\rangle _R  = \tilde c_R \frac{{(\xi  - z)}}{{lx}}\left\langle {g_{sR} \left( {0,{\bf{k}}} \right)} \right\rangle _R.
\end{equation}
The result after substituting (\ref{A1}) into (\ref{e16}) is: \\
\centerline{$g_{sR} \left( {z,{\bf{k}}} \right) \approx g_{aR} (z,{\bf{k}}) + \frac{1}{{lx}}\int\limits_z^\infty  {e^{ - \kappa _R \left( {\xi  - z} \right)} \tilde c_R \frac{{(\xi  - z)}}{{lx}}\left\langle {g_{sR} \left( {0,{\bf{k}}} \right)} \right\rangle _R d\xi }  = $}\\
\centerline{ $= g_{aR} (z,{\bf{k}}) + \tilde c_R \left\langle {g_{sR} \left( {0,{\bf{k}}} \right)} \right\rangle _R \int\limits_z^\infty  {e^{ - \kappa _R \left( {\xi  - z} \right)} \frac{{(\xi  - z)}}{{lx}} d \left[ \frac{{\left( {\xi  - z + z} \right)}}{{lx}} \right]}$.}\\
 The new variable definition $t = (\xi  - z)/\left( {lx} \right)$ gives \\
 \centerline{ $g_{sR} \left( {z,{\bf{k}}} \right) \approx g_{aR} (z,{\bf{k}}) + \tilde c_R \left\langle {g_{sR} \left( {0,{\bf{k}}} \right)} \right\rangle _R \int\limits_0^\infty  {e^{ - t} tdt}$,} \\
 where the differential $d \left[ {z/(lx)} \right] \to 0$.
As a result, $g_{sR} \left( {z,{\bf{k}}} \right) \approx g_{aR} (z,{\bf{k}}) + \tilde c_R \left\langle {g_{sR} \left( {0,{\bf{k}}} \right)} \right\rangle _R$, since $\int\limits_0^{ + \infty } {e^{ - t} t\,dt}  = 1$. And hence:
\begin{equation}
\label{A2}
g_{aR} (z,{\bf{k}}) \approx g_{sR} \left( {z,{\bf{k}}} \right) - \tilde c_R \left\langle {g_{sR} \left( {0,{\bf{k}}} \right)} \right\rangle _R.
\end{equation}
The derived (\ref{A2}) is agreed with second equation in (\ref{e12}).

Indeed, substituting (\ref{A2}) into $\frac{{\partial  g_s (z,{\bf{k}})}}{{\partial z }} =  - \kappa g_a \left( {z,{\bf{k}}} \right)$ it gives \\ \centerline{$\frac{{\partial  g_s (z,{\bf{k}})}}{{\partial z }} =  - \frac{1}{{lx}}\left[ {g_{sR} \left( {z,{\bf{k}}} \right) - \tilde c_R \left\langle {g_{sR} \left( {0,{\bf{k}}} \right)} \right\rangle _R } \right]$.}\\ It is obtained $\frac{{\partial  g_s (z,{\bf{k}})}}{{\partial z }} + \frac{1}{{lx}}g_{sR} \left( {z,{\bf{k}}} \right) = \frac{{\tilde c_R \left\langle {g_{sR} \left( {0,{\bf{k}}} \right)} \right\rangle _R }}{{lx}}$, or $\frac{{\partial  g_s (z,{\bf{k}})}}{{\partial z }} + a\,g_s(z,{\bf{k}})  = b$, where $a = \frac{1}{{lx}} \approx \kappa$ and $b = \frac{{\tilde c_R \left\langle {g_{sR} \left( {0,{\bf{k}}} \right)} \right\rangle _R }}{{lx}}$. The differential equation $\frac{{\partial  g_s (z,{\bf{k}})}}{{\partial z }} + a\,g_s(z,{\bf{k}}) = b$ has the solution in the form: $g_{sR} \left( {z,{\bf{k}}} \right) = C_1 e^{ - a\,z}  + \frac{b}{a} = C_1 e^{ - \kappa \,z}  + \tilde c_R \left\langle {g_{sR} \left( {0,{\bf{k}}} \right)} \right\rangle _R$. \\
The presented result satisfies to the conditions of the homogeneous solution (\ref{e13}). Similar finding can be derived for $\left\langle {g_{sL} \left( {\eta ,{\bf{k}}} \right)} \right\rangle _L  = \tilde c_L \kappa _L \eta \left\langle {g_{sL} \left( {0,{\bf{k}}} \right)} \right\rangle _L$ at $z < 0$ and $z \to 0$.

\section*{Appendix B}
One can consider the following integral $\lambda  = \int\limits_0^\infty \mathop {\left\langle {\frac{{\eta \kappa e^{ - \kappa \eta } }}{{lx}}} \right\rangle }\nolimits_{\theta ,\phi } d\eta $, \\ and $\kappa  = \left[ {1 - i\left( {{\bf{kl}}_{||} } \right)} \right]/l_z  = \left( {lx} \right)^{ - 1} - i k\sqrt {1 - x^2 } \cos \left( \phi  \right) /x $, since $l_z = l \cdot x$. \\
As a result: \\
$\lambda  = \int\limits_0^\infty   \mathop {\left\langle {l_z^{ - 1} \eta \kappa e^{ - \kappa \eta } } \right\rangle }\nolimits_{\theta ,\phi } d\eta  = \int\limits_0^\infty   \mathop {\left\langle {\left( {lx} \right)^{ - 2} \eta e^{ - \kappa \eta } } \right\rangle }\nolimits_{\theta ,\phi } d\eta  - k\int\limits_0^\infty   \mathop {\left\langle {lx^{ - 2} \eta \sqrt {x^2  - 1} e^{ - \kappa \eta } \cos \left( \phi  \right)} \right\rangle }\nolimits_{\theta ,\phi } d\eta$, \\ the second term is equal to zero, since the integrand is the even function with respect to $\phi$: \\
$\int\limits_0^\infty   \mathop {\left\langle {lx^{ - 2} \eta \sqrt {x^2  - 1} e^{ - \kappa \eta } \cos \left( \phi  \right)} \right\rangle }\nolimits_{\theta ,\phi } d\eta  = \frac{1}{{2\pi }}\int\limits_0^1 {dx} \int\limits_0^\infty  {d\eta } \int\limits_0^{2\pi } {lx^{ - 2} \eta \sqrt {x^2  - 1} e^{ - \kappa \eta } \cos \left( \phi  \right)\,} d\phi  = 0$. \\
It corresponds to the approximation $\kappa  = \left( {lx} \right)^{ - 1}$.
Finally the first term has analytical solution as follows:
\\
$\int\limits_0^\infty   \mathop {\left\langle {\left( {lx} \right)^{ - 2} \eta e^{ - \kappa \eta } } \right\rangle }\nolimits_{\theta ,\phi } d\eta  = \frac{1}{{2\pi }}\int\limits_0^1 {dx} \int\limits_0^\infty  {d\eta \left( {lx} \right)^{ - 2} \eta e^{ - {\eta  \mathord{\left/
 {\vphantom {\eta  {\left( {lx} \right)}}} \right.
 \kern-\nulldelimiterspace} {\left( {lx} \right)}}} } \int\limits_0^{2\pi } {d\phi } e^{\left[ {i \cdot k\eta \sqrt {1/x^2  - 1} \cos \left( \phi  \right)} \right]}  = $ \\
 $ = \int\limits_0^1 {dx} \int\limits_0^\infty  {d\eta \left( {l\,x} \right)^{ - 2} \eta \,e^{ - {\eta  \mathord{\left/
 {\vphantom {\eta  {\left( {l\,x} \right)}}} \right.
 \kern-\nulldelimiterspace} {\left( {l\,x} \right)}}} } J_0 \left( {k\eta \sqrt {1 - x^2 } /x} \right) = \int\limits_0^1 {dx} \left[ {1 + k^2 l^2 \left( {1 - x^2 } \right)} \right]^{ - 3/2}  = \left[ {1 + k^2 l^2 } \right]^{ - 1}$.

\end{document}